\documentclass[prd,floats,superscriptaddress,eqsecnum,floatfix,
showpacs,12pt]{revtex4}

\usepackage{amssymb,amsmath}
\usepackage{epsfig}

\pdfoutput=1
\usepackage{amssymb,amsmath}
\usepackage{epsfig}
\usepackage{color}
\usepackage{datetime}

\newcommand{\vp}{\varphi}

\newcommand{\be}{\begin{equation}}
\newcommand{\ee}{\end{equation}}
\newcommand{\ba}{\begin{eqnarray}}
\newcommand{\ea}{\end{eqnarray}}
\newcommand{\beg}{\begin{gather*}}
\newcommand{\eng}{\end{gather*}}
\newcommand{\sgn}{\,\mbox{sgn}}
\newcommand{\erf}{\,\mbox{erf}}

\newcommand{\hh}{,\hspace{0.5cm}}

\newcommand{\eq}[1]{(\ref{#1})}

\newcommand{\lap}{\bigtriangleup}

\newcommand{\n}[1]{\label{#1}}

\newcommand{\ins}[1]{{\mbox{\tiny #1}}}

\def\Xint#1{\mathchoice
{\XXint\displaystyle\textstyle{#1}}%
{\XXint\textstyle\scriptstyle{#1}}%
{\XXint\scriptstyle\scriptscriptstyle{#1}}%
{\XXint\scriptscriptstyle\scriptscriptstyle{#1}}%
\!\int}
\def\XXint#1#2#3{{\setbox0=\hbox{$#1{#2#3}{\int}$ }
\vcenter{\hbox{$#2#3$ }}\kern-.6\wd0}}

\def\pvint{\Xint-}

\begin{document}

\title{Radiation from an emitter in the ghost free scalar theory}

\author{Valeri P. Frolov\thanks{E-mail:
vfrolov@ualberta.ca}\, and
Andrei Zelnikov\thanks{E-mail: zelnikov@ualberta.ca}\\
Theoretical Physics Institute, Department of Physics\\
University of Alberta, Edmonton, AB, Canada T6G 2E1}

\today, \currenttime

\begin{abstract}
We study radiation emitted by a time-dependent source of a scalar massless field in the framework of the ghost-free modifications of the theory. We consider a simple model of the emitter: namely,we assume that it is point-like and monochromatic. We focused on the most common versions of the ghost-free theory, where the propagator $\Box^{-1}$ is modified as follows $\exp(-(\Box/\mu^2)^N) \Box^{-1}$, where $\mu$ is the characteristic mass-scale of such $GF_N$-theory. We demonstrated that far from the source, in the wave-zone, the radiation asymptotically converges to its "classical" value for any $N\ge 1$. However, in the near-zone the behavior of the field is quite different from the "classical" case. The difference of field amplitude for the ghost-free field and for the classical one  has an oscillatory behavior in this domain. A number of oscillations increases with $N$. The amplitude of these oscillations remain finite for even $N$, while it infinitely grows with frequency for odd $N$. This behavior
indicates that even in the classical domain $GF_N$ theories might have pathological behavior.
\end{abstract}


\maketitle


\section{Introduction}

A well known problem of the General Relativity is its ultra-violet (UV) incompleteness. In the classical theory it manifests itself as existence of singularities (see e.g. \cite{Hawking:1973uf}). In a quantum regime the General Relativity is non-renormalizable. It is expected that both of the problems can be "cured" by admitting higher in curvature terms the the gravitational action \cite{Tomboulis:1997gg, Modesto:2011kw, Talaganis:2014ida, Tomboulis:2015gfa, Tomboulis:2015esa}. However, such higher-derivative theories inevitably has a common severe problem: they contain ghosts \cite{Stelle:1976gc}. In a general
case, this additional unphysical degrees of freedom with negative energy result
in classical and quantum instability of the theory \cite{Stelle:1977ry,Barnaby:2007ve,Barnaby:2010kx}.
The problem of ghosts can be solved if one allows an infinite number of
derivatives in the action of the field, that makes it non-local. This can be
achieved by a special modification of the standard propagator. Suppose one
considers a scalar theory with the propagator $(\Box-m^2)^{-1}$ and modifies it
as follows $[a(\Box) (\Box-m^2)]^{-1}$, where $a(z)$ is a entire function of the
complex variable $z$, which does not have poles in the complex plane of $z$. For
such a modification the propagator has only one pole, which describes a particle
with mass $m$. Such ghost-free theories were widely discussed mainly in connection of the well
known problem of singularities in gravity (see, e.g.
\cite{Asorey:1996hz, Biswas:2011ar, Modesto:2012ys, Biswas:2013cha, Biswas:2013kla, Modesto:2014lga, Shapiro:2015uxa}). Their application to the problems of cosmology \cite{Biswas:2005qr, Biswas:2010zk, Calcagni:2013vra, Biswas:2012bp} and
black holes can be found in \cite{Nicolini:2005zi, Hossenfelder:2009fc, Modesto:2010uh, Bambi:2013gva, Chialva:2014rla, Conroy:2015wfa, Li:2015bqa}.

The simplest choice of the form-factor $a(z)$, which is usually adopted in the publications, is
\be\n{az}
a(z)=\exp[(z/\mu^2)^N]\, .
\ee
We shall use the notation $GF_N$ for such a theory. The mass $\mu$ determines the energy scale where ghost-free modification of the theory becomes important. In the limit $\mu\to\infty$ one has $a(0)=1$, so that the theory reproduces correctly the results of the standard theory.

There are several papers discussing the gravitational field of a point-mass in the linearized version of the ghost-free gravity \cite{Gruppuso:2005yw,Conroy:2014eja, Modesto:2014eta, Frolov:2015bia, Frolov:2015bta, Frolov:2015usa}. The main conclusion is that the gravitational potential in the ghost-free modified theory is finite and regular at the location of the point mass, while at far distance it converges to the standard Newtonian potential. This result is valid for any value of $N$. At the same time one can expect that theories with odd value of $N$ may have "bad" behavior for time-dependent sources. The reason for this is the following. In a static case $\Box$ reduces to the Laplace operator $\lap$. The operator $-\lap$  is non-negative definite as well as any power of it. In the general time-dependent case, the $\Box$ operator has both positive and negative eigen-values, which for odd $N$ may lead to instability. The purpose of this paper is to analyze behavior of radiation emitted by a time-
dependent source in  the ghost-free theory with different values of $N$.


\section{Scalar field from a monopole emitter}

We restrict ourselves by considering a ghost-free scalar field in four dimensional spacetime and use a simple model of a monopole point-like emitter. These restrictions play mainly technical role
and the obtained results can be generalized to the case of the ghost-free fields with non-zero spin in higher dimensions. Similarly, the model of the emitter can be modified. Our starting point is the action
\be\n{f2}
S=\int dx|g|^{1/2} \left({1\over 8\pi}\varphi a(\Box)(\Box-m^2)\varphi
+j\varphi\right)\, .
\ee
where the entire function $a(z)$ is given by \eq{az}. The corresponding field equation reads
\be\n{f4}
a(\Box)(\Box-m^2)\varphi =-4\pi j\, .
\ee
For simplicity we consider a massless case $m=0$.
We assume that the source of the field $\vp$ is point-like and its scalar charge oscillates with the frequency $\Omega$. The equation \eq{f4} is linear and its coefficients are real. It is convenient to choose its source $j$ in the form $j=\mbox{Re}\,\hat{j}$, where
\be\n{jj}
\hat{j}(x)=q \,e^{i\Omega t}\delta(\mathbf{x})\hh x=(t,\mathbf{x})\, .
\ee
After performing the calculations and getting the complex field $\hat{\vp}$ created by this complex source (\ref{jj}) one can obtain the value of the field $\vp$ by taking the real part of this complex answer.
The complex scalar field $\hat{\vp}$ created by the charge can be expressed in terms of the
retarded Green function $G_\ins{Ret}(x,x')$
\be
\hat{\vp}(x)=4\pi\int dx'\,G_\ins{Ret}(x,x')\hat{j}(x').
\ee
In momentum representation one has
\be
\hat{\varphi}(t,\mathbf{x})={q\over 2\pi^2}\int
d\omega\,d\mathbf{p}\,e^{-i\omega t+i\mathbf{px}}
G_\ins{Ret}(\omega,\mathbf{p})\,\delta(\omega+\Omega),
\ee
\be
G_\ins{Ret}(x,x')=\int {d\omega\,d\mathbf{p}\over (2\pi)^4}e^{-i\omega
(t-t')+i\mathbf {p(x-x')}}G_\ins{Ret}(\omega,\mathbf{p}).
\ee
In  the case of $GF_N$ theory the retarded Green function can be written as
\ba
&&G_\ins{Ret}(\omega,\mathbf{p})=G_\ins{Ret}(\omega,p)
={\exp{\left[-\left({p^2-\omega^2\over \mu^2}\right)^N\right]}
\over(\omega+i\epsilon)^2-p^2}\hh p=\sqrt{\mathbf{p}^2}\, ,\\
&&={\cal P}{1\over\omega^2-p^2}\,e^{-\left({p^2-\omega^2\over \mu^2}\right)^N}
-i\pi\sgn(\omega)\delta(\omega^2-p^2)\, .
\ea

Then, after integration over angles, we get ($r=\sqrt{\mathbf{x}^2}$)
\ba
&&\hat{\varphi}(t,\mathbf{x})={2q\over \pi r}\int_0^{\infty}
dp\,p \sin(pr)\,e^{i\Omega t}\, G_\ins{Ret}(-\Omega,p)\n{pp1}\\
&&={q \,e^{i\Omega t}\over r}\left[{2\over \pi}\,\pvint_{0}^{\infty}
dp\,{p\sin(pr)\over \Omega^2-p^2}e^{-\left({p^2-\Omega^2\over
\mu^2}\right)^N}
+i\,\sin(\Omega r)\right]\n{pp2}\\
&&={q \,e^{i\Omega t}\over r}\left[{1\over \pi}\,\pvint_{-\infty}^{\infty}
dp\,{\sin(pr)\over \Omega-p}e^{-\left({p^2-\Omega^2\over
\mu^2}\right)^N}
+i\,\sin(\Omega r)\right].
\n{pp3}
\ea
To obtain the last relation we used the fact that the integrand in (\ref{pp2}) is an even function of $p$. In the last two lines we used a notation $\pvint$ to indicate that the corresponding integral should be calculated according to the principal value prescription. In the limit $\mu=\infty$ the $GF_N$ theories reduce to the ordinary massless scalar theory, satisfying the field equation
\be
\Box \hat{\varphi}_0(t,\mathbf{x})=-4\pi \hat{j}.
\ee
In this case the principal value integral (see \eq{pp3}) has exactly the form of
the Hilbert transform of $\sin(pr)$ (see \cite{King:2009} table 1.5 Eq.(5.1))
\be
{1\over \pi}\pvint_{-\infty}^{\infty}
dp\,{\sin(pr)\over \Omega-p}=-\cos(\Omega r).
\ee
Thus, the result of the integration becomes
\be
\hat{\varphi}_0(t,\mathbf{x})={q \,e^{i\Omega t}\over  r}\left[{1\over\pi}\,\pvint_{-\infty}^{\infty}
dp\,{\sin(pr)\over \Omega-p}
+i\,\sin(\Omega r)\right]=-{q\,e^{i\Omega (t-r)}\over  r}.
\ee
Let us consider the difference of the solution for the scalar field $\hat{\varphi}(t,\mathbf{x})$  created by the source \eq{jj} in $GF_N$ theory and in the ordinary massless scalar theory
\be\label{psi0}
\hat{\psi}(t,\mathbf{x})=\hat{\varphi}(t,\mathbf{x})-\hat{\varphi}_0(t,\mathbf{x}),
\ee
\be\label{psi}
\hat{\psi}(t,\mathbf{x})={q \,e^{i\Omega t}\over r}h(\Omega,r),
\ee
where
\be\label{f}
h(\Omega,r)={2\over \pi}\int_{0}^{\infty}
dp\,\sin(pr)\,f(p)
\hh
f(p)={p\over \Omega^2-p^2}\left[e^{-\left({p^2-\Omega^2\over
\mu^2}\right)^N}-1\right].
\ee
For all $N\ge 1$ the function $f(p)$ is finite at $p=\Omega$ and, hence, the principal value integral reduces to the ordinary one.
$f(p)\in C^{\infty}$ is a smooth function on the interval $[-\infty,\infty]$ decreasing at infinity as $p^{-1}$ and  satisfying the condition $f(-p)=-f(p)$. Using these properties one can estimate an asymptotic of the function $h(\Omega,r)$ at large $r$. Note that this function has the form of the Fourier transform of  $f(p)$
\be
h(\Omega,r)={2\over \pi}\mbox{Im}\int_{0}^{\infty}
dp\,\exp(ipr)\,f(p).
\ee
One can show (see, e.g., \cite{Fedoryuk:1989} page 91, Eq.(1.10)) that at large
distances $r\gg \mu^{-1}$
\be\label{h1}
h(\Omega,r)=\sum_{l=0}^n (-1)^{l+1} r^{-2l-1}{d^{2l}\over dp^{2l}}
f(p)\Big|_{p=0}+O(r^{-2n-1}).
\ee
Because  of the symmetry property $f(-p)=-f(p)$ all even derivatives of $\partial^{2l}_p f(p)$ vanish at $p=0$. As the result of \eq{h1} and  this property,  $h(\Omega,r)$ falls off at large $r$  faster than any power of $1/r$.
This means that for all $GF_N$ theories the nonlocal effects become important only at distances from the charge of the order of $\mu^{-1}$. At large distances $r\gg \mu^{-1}$ only the usual asymptotic
\be
\hat{\varphi}(t,\mathbf{x})\sim \hat{\varphi}_0(t,\mathbf{x})=-{q\,e^{i\Omega (t-r)}\over  r}\hh
\varphi(t,\mathbf{x})=\mbox{Re} (\hat{\varphi}(t,\mathbf{x}))
\ee
survives.

At $r=0$ the scalar field $\varphi$ is finite for all $GF_N$ theories with $N\ge 1$. It has the form
\be\label{r0}
\hat{\varphi}(t,0)=q\,e^{i\Omega t}\eta(\Omega)\hh
\eta(\Omega)=i\Omega+{2\over \pi}\pvint_{0}^{\infty}
dp\,{p^2\over \Omega^2-p^2}e^{-\left({p^2-\Omega^2\over
\mu^2}\right)^N}.
\ee


\section{Nonlocal effects in $GF_N$ scalar theory}

\begin{figure}[tbp]
\centering
\includegraphics[width=8cm]{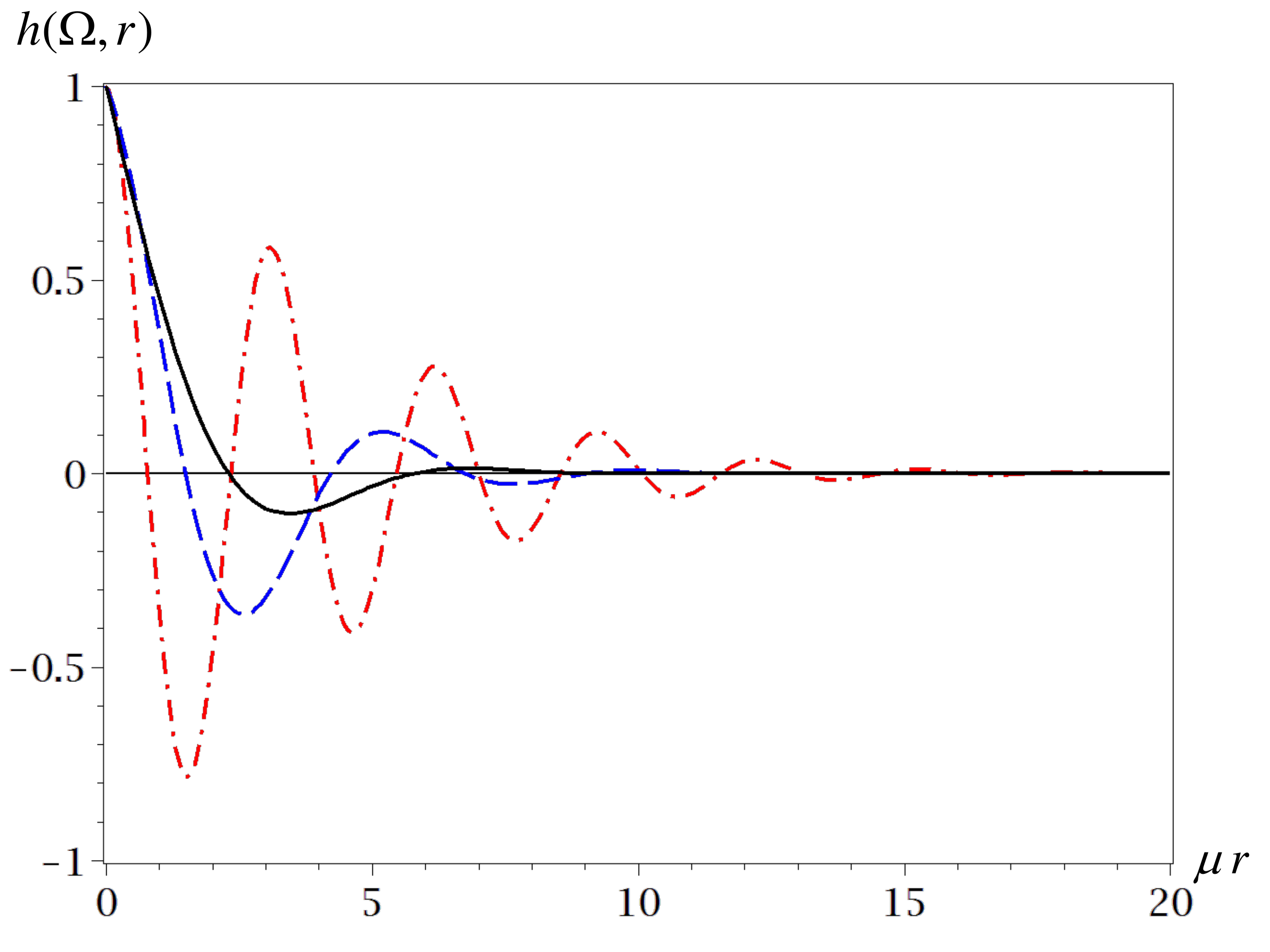}\hfill
\caption{$h(\Omega,r)$ in $GF_2$ theory as the function of $r$ for $\Omega/\mu=0$ (solid line), $\Omega/\mu=1$ (dashed line), and $\Omega/\mu=2$ (dash-dotted line).}
\label{Fig_1}
\end{figure}

\begin{figure}[tbp]
\centering
\includegraphics[width=8cm]{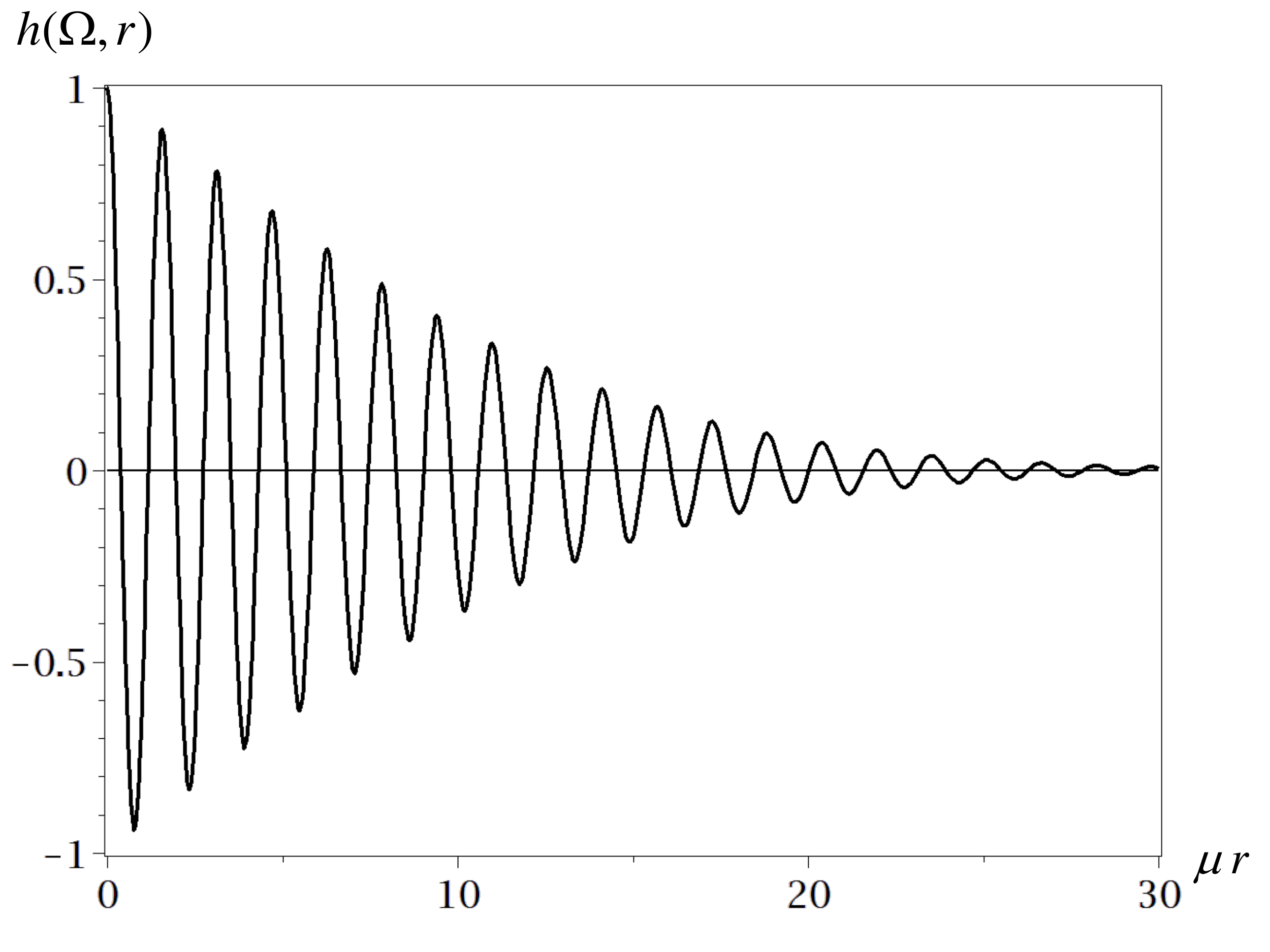}\hfill
\caption{$h(\Omega,r)$ in $GF_2$ theory  as the function of $r$ for $\Omega/\mu=4$.}
\label{Fig_2}
\end{figure}

\begin{figure}[tbp]
\centering
\includegraphics[width=8cm]{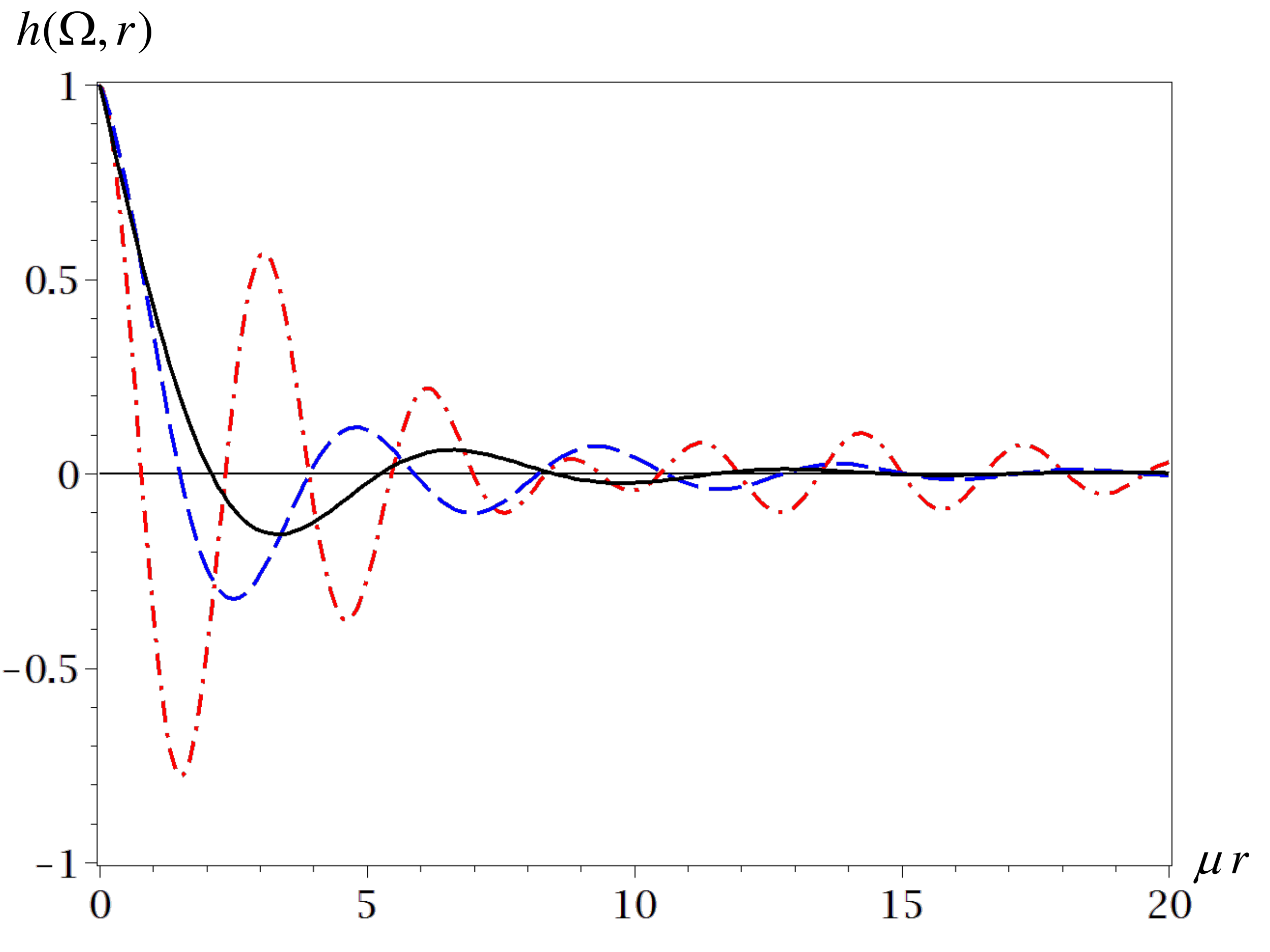}\hfill
\caption{$h(\Omega,r)$ in $GF_4$ theory as the function of $r$ for $\Omega/\mu=0$ (solid line), $\Omega/\mu=1$ (dashed line), and $\Omega/\mu=2$ (dash-dotted line).}
\label{Fig_3}
\end{figure}

Consider an example of $GF_2$ scalar theory in more detail. The ghost-free nonlocal contributions to the massless scalar theory are described by the field $\psi$. It describes standing waves with the amplitude at large radii falling off faster than any power of $1/r$. The field $\psi$ (see \eq{psi}-\eq{f}) is proportional to the dimensionless function
$h(\Omega,r)$
which depends on the dimensionless combinations $\Omega/\mu$ and $\mu r$ only.
For various values of $\Omega$ the dependence of $h(\Omega,r)$ on the radius is
depicted in Figs.\ref{Fig_1}-\ref{Fig_2}. The potential
created by a point source \eq{jj} in the ghost-free models is
\be\label{psi0}
\hat{\varphi}(t,\mathbf{x})=-{q \,e^{i\Omega t}\over
r}\left[e^{-i\Omega r}-h(\Omega,r)\right].
\ee
Notice that in ghost-free
models at small $r$ the function $h(\Omega,r)=1+O(r)$ and the
potential is finite at $r=0$ (see \eq{r0}).
All $GF_N$ theories with even $N$ look alike. For example, $h(\Omega,r)$ in the
case of $GF_4$ theory is depicted in Fig.\ref{Fig_3}

\begin{figure}[tbp]
\centering
\includegraphics[width=8cm]{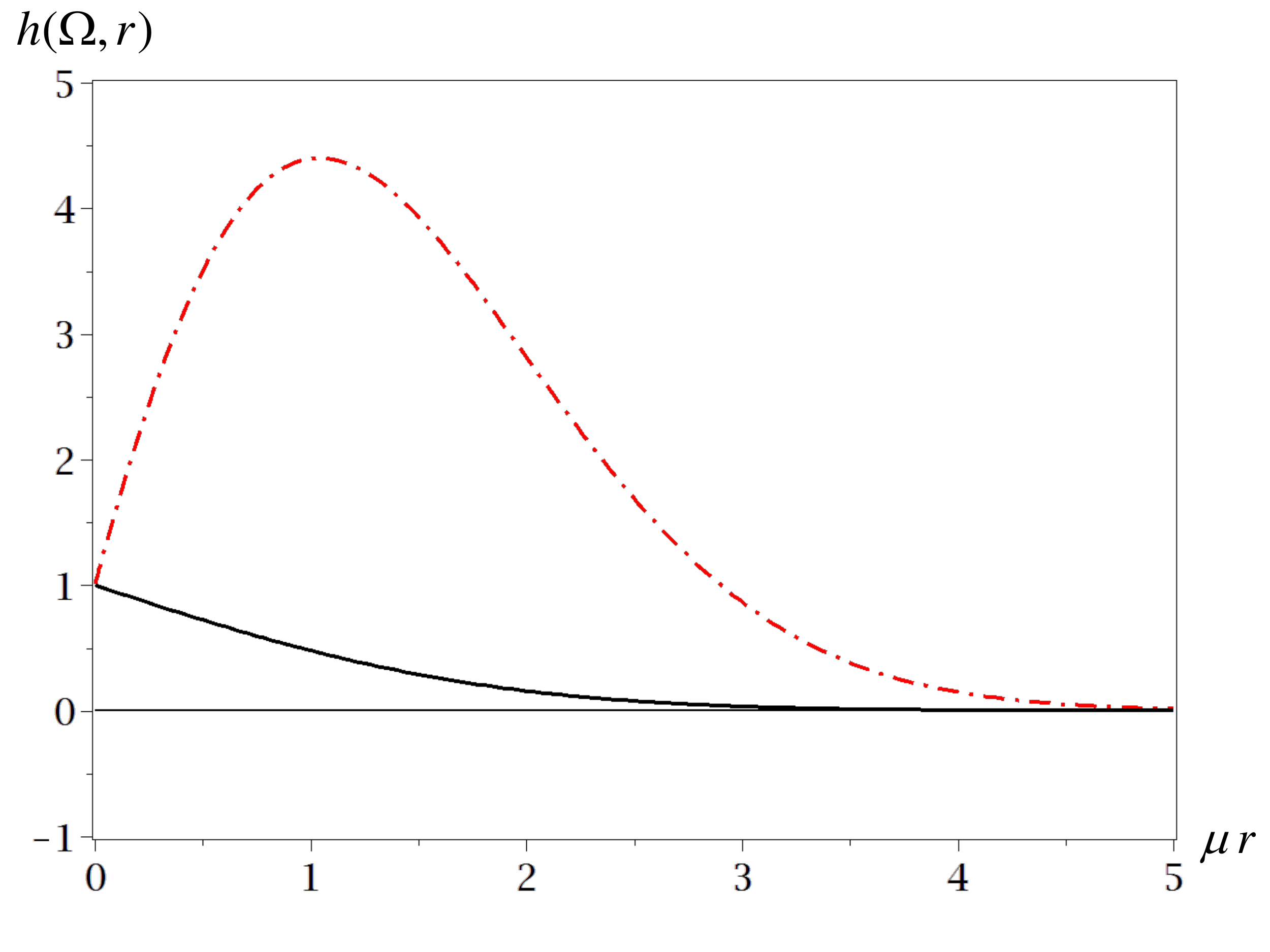}\hfill
\caption{$h(\Omega,r)$ in $GF_1$ theory  as the function of $r$ for  $\Omega/\mu=0$ (solid line) and  $\Omega/\mu=2$ (dash-dotted line)}
\label{Fig_4}
\end{figure}

Now let us consider $GF_N$ theories of the odd order $N$. They also look
similar to each other. A typical example  is $GF_1$ scalar theory. It reveals
the key properties of the whole class of theories of the odd order. In this
particular case one can compute the field of the oscillating monopole charge analytically.
The function \eq{f} takes the form of the Hilbert transform (see
\cite{King:2009})
\be\label{h}
h(\Omega,r)={1\over \pi}\pvint_{-\infty}^{\infty}
{dp \over \Omega-p}\sin(pr)\left[e^{-{p^2-\Omega^2\over
\mu^2}}-1\right]
\ee
and can be computed explicitly (see \cite{King:2009} table 1.5
Eqs.(5.1),(5.138))
\be
h(\Omega,r)=\cos(\Omega r)-\mbox{Re}\left[e^{i\Omega r}\erf\left({\mu r\over
2}+i{\Omega\over\mu}\right)\right].
\ee
For  small frequencies $\Omega\ll\mu$ this function behaves similarly to that in the even $N$ theories. The crucial difference appears at $\Omega>\mu$ (see Fig.\ref{Fig_4}).
Comparing this plot with Figs.\ref{Fig_1}-\ref{Fig_2} one can see that, while in
all even $N$ amplitude the nonlocal correction $h(\Omega,r)$ is always within
the interval $[-1,1]$, in the  $GF_1$ theory near $r\sim\mu^{-1}$ it
exponentially grows with growing $\Omega$. This behavior can be derived from the
asymptotic of the error function $\erf$ of a complex argument. At
$\Omega\gg\mu^2 r$ we can use the asymptotic expansion of the
$\erf$ function for a complex argument $z$ (see, e.g., \cite{Temme:2010} Eq.
(7.12.1))
\be
\erf(z)=1+{1\over \sqrt{\pi} z}e^{-z^2}(1+{1\over 2z^2}+\dots)\hh
|z|\to\infty\hh |\mbox{arg} z|<{3\pi\over 4}.
\ee
Then we obtain
\be
h(\Omega,r)={\mu^3 r\over
2\sqrt{\pi}\,\Omega^2}e^{\Omega^2/\mu^2}\left(1+O(\Omega^{-2})\right).
\ee
Thus, at any finite $r\sim\mu^{-1}$ and large frequencies, the nonlocal corrections
to the ordinary scalar field become exponentially large $\sim \exp(\Omega^2/\mu^2)$.
This is a reflection of a strong instability for the field of time-dependent sources
in $GF_1$ theory. This instability is an inherent property of all $GF_N$ with odd
$N$.

\section{Discussion}

Ghost-free models have been intensively discussed recently. The ghost-free modification of the gravity improves its UV behavior both in the classical and quantum domains. This open an interesting possibility that the long-standing problems of the General relativity, its black-hole and cosmological singularities, can be resolve in the framework of the properly chosen ghost-free model. In this paper we study some properties of the ghost-free theories with time-dependent sources of the field. We made a number of simplifying assumptions. Instead of the gravity, we consider a scalar massless field, and we choose a simple model of the time-dependent source: a point-like emitter of the monochromatic radiation. We examined so called $GF_N$ models of the ghost-free theory. We obtained solutions for the scalar field in such a model and demonstrated that at far distance, in the wave zone, this field asymptotically converge to its "classical" value. This happens for any $N\ge 1$. However, the structure of the field
in the near zone is qualitatively different for odd and even values of $N$. Namely, the field remains bounded for even $N$, while for odd $N$ its amplitude infinitely grows with the frequency of the emitter. This result supports a conclusion that all $GF_N$ theories with odd $N$ have a potential problem: their solutions may be unstable. This means that one can expect similar pathology of such theories in a more general set-up.


\section*{Acknowledgments}

The authors thank the Natural Sciences and Engineering Research Council of Canada and the Killam Trust for their financial support. A part of this work was done during V.F.'s stay at the University of Oldenburg. He gratefully acknowledge support by the DFG Research Training Group 1620 “Models of Gravity” and thanks Prof. Jutta Kunz for her kind hospitality.



\end{document}